# MEASUREMENT NOISE FLOOR FOR A LONG-DISTANCE OPTICAL CARRIER TRANSMISSION VIA FIBER

G. GROSCHE[1*], O. TERRA[1], K. PREDEHL[1,2], T. HÄNSCH[2], R. HOLZWARTH[2],
B. LIPPHARDT[1], F. VOGT[1], U. STERR[1], H. SCHNATZ[1]

[1]*Physikalisch-Technische Bundesanstalt, Bundesallee 100,
D-38116 Braunschweig, Germany*

[2]*Max-Planck-Insitut für Quantenoptik, Hans-Kopfermann-Str. 1
D-85748 Garching, Germany*

* E-mail: Gesine.Grosche@ptb.de

We investigated the measurement floor and link stability for the transfer of an ultra-stable optical frequency via an optical fiber link. We achieved a near-delay-limited instability of $3 \times 10^{-15}/(\tau \cdot \text{Hz})$ for 147 km deployed fiber, and $10^{-20}$ ($\tau = 4000$ s) for the noise floor.

## 1. Introduction

Ultra-precise optical frequency standards (OFS) have reached a relative instability and accuracy below $10^{-16}$ [1], putting increased demands on methods for comparing standards, which are usually located in different laboratories. Long-distance transmission of either a modulated carrier [2] or an ultra-stable optical carrier frequency near 200 THz [3,4,5] over tens or hundreds of kilometers using standard telecommunication fiber is a promising, relatively new alternative to satellite-based methods. It allows remote comparison of OFS and may support precision experiments such as a search for variations of the fine-structure constant.

As part of a German research fiber network, we have set up a fiber link with two telecommunication fibers running in parallel which bridge almost 70 km distance between PTB and a laboratory of the Institute of Quantum Optics at the University of Hanover. At Hanover, a frequency standard based on neutral magnesium is located: a recent frequency measurement [6] was limited by the transportable microwave frequency reference (Cs clock) used. Thus, the aim of the fiber link is to deliver a precisely known and highly stable optical frequency to the remote optics lab, for a more precise frequency measurement of the intercombination transition of neutral magnesium and assessment of laser sources used in the spectroscopy set-up.

### 1.1. *Measurement noise floor and first measurements on 147 km link.*

We report two results [7]: to achieve optimal long-term stability of stabilized fiber links, we have investigated noise from fiber leads and components included in the interferometer set-up which is used stabilize the fiber link, and optimized the interferometer. Secondly, we report a stabilized fiber link of 147 km length, consisting of two underground 73 km-fibers put in series, where each fiber connects PTB with Hanover. We also outline future applications.

## 2. Method and experimental realization

### 2.1. *Laser stabilization and frequency reference*

As light source, we stabilize a 50 mW fiber laser at 1542 nm (Koheras Adjustik, free-running fast line-width ~ 2kHz) to a master laser, which is a cavity stabilized optical clock laser at 657 nm with a line-width of about 1 Hz. The clock laser is part of the Ca neutral atom frequency standard at PTB, and a fiber based fs-laser frequency comb acts as transfer oscillator to obtain a phase-lock between the two lasers, bridging > 200 THz [8]. From earlier independent measurements of the residual phase noise [8] we estimate a resulting coherence length of the stabilized fiber laser of more than 100 000 km; the contribution of laser noise to detected fiber link noise is therefore negligible.

### 2.2. *Fiber link and compensation interferometer set-up*

The fiber link and compensation interferometer are shown in fig. 1. Two parallel fiber links, F1 and F2 (in the same cable) connect our lab at PTB with the computer centre of the University of Hanover (RZ-H). Fibers F1 and F2 were connected with a short patch-cord at RZ-H, resulting in a fiber link starting and ending at PTB and going via RZ-H. The single-pass link length is approximately 147 km and incorporates one bi-directional erbium doped fiber amplifier. More recently, F1 and F2 continue from RZ-H over a distance of less than 1 km to the Mg optical frequency standard.

Perturbations leading to optical path length fluctuations appear as a Doppler shift of the light transmitted through the fiber link. These fluctuations are compensated with a scheme based on the well-established method described in [9]: detecting a double pass (return) signal and applying a correction signal derived from the return signal. As observed in our first experiments on a 86 km fiber link in Paris [3] and also reported e.g. by Williams [4], the long-term stability of a fiber link is limited by residual noise of the interferometer used to detect the fiber link noise. We have developed and tested different designs for the interferometer, using off-the-shelf fiber-optic components and minimizing out-of-loop fiber leads in a compact, portable set-up.

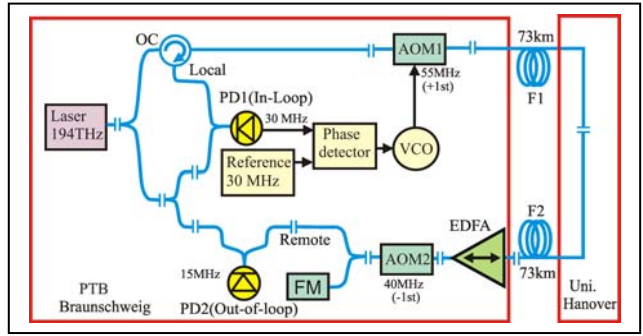

Fig. 1: Schematic set-up for active fiber noise compensation of 147 km fiber link. Laser: fiber laser stabilized to optical clock laser via fs-comb [8], OC: optical circulator, AOM: Acousto-optic modulator, VCO: voltage-controlled oscillator, PD1: in-loop photo-detector, PD2: out-of-loop photo-detector, EDFA: Erbium-doped fiber amplifier, FM: Faraday mirror.

## 3. Results for interferometer noise floor and 147 km fiber link

We detect beat signals between out-going laser light and returned light (In-Loop signal), and remote light (Out-of-loop signal). Using zero dead-time totalizing counters acting as $\Pi$-estimators [5,10], we calculate the Allan standard deviation (ADEV) as measure of frequency instability.

### 3.1. *Optimizing the compensation interferometer*

Unstabilized optical fiber in the lab with lengths ~ 1 m… 20 m showed phase noise from $10^{-3}$… $10^{-1}$ rad$^2$/Hz x f$^{-2}$, and an ADEV of up to $2 \times 10^{-15}$ for integration times up to 100 s. Balancing the fiber length in the interferometer to minimize uncompensated fiber, we obtain a measurement noise floor around $10^{-7}$ rad$^2$/Hz x f$^{-2}$, or (with recording bandwidth > 10 kHz) $3 \times 10^{-17}$ at 1s for the ADEV, reaching $10^{-20}$ after 1 hour, see green open triangles in fig.2.

### 3.2. *Realization of stabilized 147 km fiber link over deployed fiber*

Fig. 2 shows the relative instability of the remote signal when stabilizing the 147 km fiber link using the return signal (after 294 km), see full black squares. The accuracy of the frequency transfer was found to be within the ADEV value for the same integration time, reaching $10^{-18}$ after one hour. Phase-noise analysis (fig. 3) shows that our results are close to the fundamental delay-time limit [4].

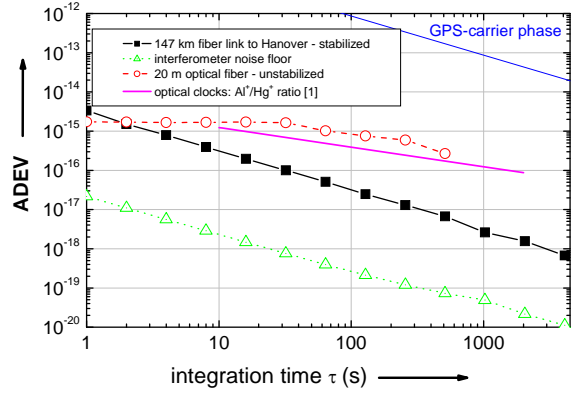

Fig. 2: Relative frequency instability given by Allan standard deviation (ADEV) using Π-type counters with a recording bandwidth > 10 kHz. Green triangles: interferometer noise floor (obtained by shorting the fiber link and detecting the stabilized remote signal); full black squares: remote signal of 147 km fiber link, stabilized. Open red circles: remote signal after 20 m of unstabilized fiber. For comparison: reported data for $Al^+/Hg^+$ clock comparison & for satellite based frequency transfer (GPS carrier phase)

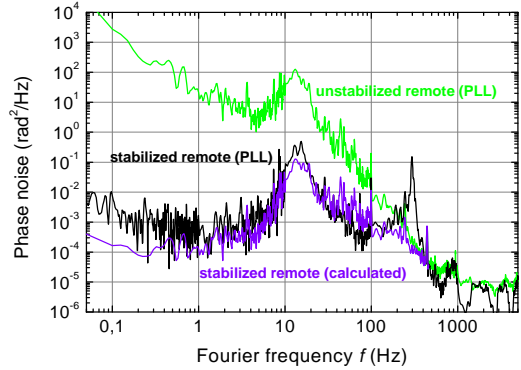

Fig. 3: Phase noise density ($rad^2$/Hz) for beat signal detected at remote end (after 147 km). Green line: without stabilization; black line with fiber stabilization; purple line: theoretical limit due to "delay-unsuppressed noise" [4]: comparison of black and purple line show close-to-optimal noise suppression.

## 4. Implication for remote frequency measurement, future extensions of fiber link and short distance frequency comparisons

We have characterized a fiber loop of 147 km length, consisting of two 73 km-fibers in series, where each fiber connects PTB with Hanover. We obtained a relative instability and uncertainty of approximately $3 \times 10^{-15}/\tau$. This is significantly lower than the expected relative instability and uncertainty for the Mg-frequency standard; we are currently preparing a frequency measurement and laser characterization at university of Hanover via one 73 km fiber. We also aim to extend the current fiber network to a link spanning 900 km to the MPQ in Garching/Munich. Finally, our improved fiber interferometer noise floor will allow very low instability measurements (ADEV $< 10^{-16}/\tau\cdot$Hz) up to a day over distances up to a few kilometers.

## 5. Future applications of optical carrier transmission via fiber

The transfer of a known, ultra-stable frequency with a relative instability in the $10^{-14}$ to $10^{-19}$ range over distances up to hundreds of kilometers between any "local" and "remote" laboratory connected by optical fiber makes precision frequency metrology available to a new "class" of end users. With commercially available fs-frequency combs, the delivered optical frequency may be mapped to any optical [8] or microwave [11,12] frequency in the remote lab.

Thus fiber-based frequency transmission can be used for the remote frequency measurement of optical [13] or rf-transitions, the remote generation of precision microwave signals, remote characterization of laser sources and remote spectroscopy and sensing.

**Acknowledgments**

This work was supported by SFB 407 / Deutsche Forschungsgemeinschaft.

**References**

1. T. Rosenband *et al*., *Science* 28 March 2008, **319** (5871), p. 1808
2. O. Lopez, A. Amy-Klein, C. Daussy *et al.*, *Eur. Phys. J. D* **48**, 35 (2008)
3. G. Grosche, B. Lipphardt, H. Schnatz, G. Santarelli *et al*., in: OSA Technical Digest, CLEO-USA, paper CMKK1 (2007)
4. N.R. Newbury, P.A. Williams, W.C. Swann, *Opt. Lett*. **32**, 3056 (2007); P.A. Williams, W.C. Swann and N.R. Newbury, *JOSA B* **25,** 1284 (2008)
5. H. Jiang, F. Kefelian *et al.*, arXiv:0807.1882; *JOSA B* **25,** 2029 (2008)
6. J. Friebe, A. Pape, M. Riedmann *et al.*, *Phys. Rev. A* **78**, 033830 (2008)
7. O. Terra, G. Grosche, K. Predehl *et al.*, in preparation
8. G. Grosche, B. Lipphardt and H. Schnatz, *Eur. Phys. J. D* **48**, 27 (2008)
9. L.S. Ma, P. Jungner, J. Ye and J L. Hall, *Opt. Lett.* **19**, 1777 (1994)
10. S.T. Dawkins, J.J. McFerran, A.N. Luiten, *IEEE Tr. UFFC* **54**, 918 (2007)
11. A. Bartels, S.A. Diddams, C.W. Oates *et al*., *Opt. Lett*. **30**, 667 (2005)
12. B. Lipphardt *et al.,* arXiv:0809.2150 (2008) accepted *IEEE Tr. Inst. Meas*.
13. F.-L. Hong *et al.*, arXiv:0811.1816v1 (2008)